# Design optimization of flux concentrators for magnetic tunnel junctions-based sensors

Thomas Brun[1,2], Javier Rial[1], Lucia Risoli[1], Johanna Fischer[1], Philippe Sabon[1], Guillaume Jannet[2], Matthieu Kretzschmar[2], Hélène Béa[1,3,*], Claire Baraduc[1]

[1]Univ. Grenoble Alpes, CEA, CNRS, Grenoble INP, SPINTEC, 17, avenue des Martyrs, 38054 Grenoble, France
[2]LPC2E, UMR7328 CNRS and Université d'Orléans, 3 avenue de la recherche scientifique, Orléans, France
[3]Institut Universitaire de France (IUF)

* Corresponding author: Helene.bea@cea.fr

Abstract

Miniaturized, ultra-sensitive and easily integrable magnetometers are needed for many applications, like space exploration or health monitoring. Achieving this goal requires a magnetic sensor with high sensitivity and low noise. High sensitivity (>1000 %/mT) can be obtained by integrating high gain permalloy flux concentrators (FC). And reducing the magnetic 1/f noise can be realized by increasing the number of magnetic tunnel junctions (MTJs) in the air-gap of the FC. However, this is obtained at the expense of a wider air-gap and consequently a decrease of the magnetic gain and thus of the sensitivity. In this paper, we explore a design optimization scheme in order to find the best trade-off between high FC gain and low magnetic noise. To model the gain of the flux concentrator, we propose two complementary approaches; one is based on finite elements simulations of the FC gain where the influence of geometrical parameters of the air-gap is investigated. Then, in a second step, we propose an analytical formula consistent with all our simulations results and based on magnetic reluctance. Finally, we derive an analytical model of the sensor detectivity from which we can extract the optimal sensor design which allows an improvement by three orders of magnitude of the performances compared to a single junction.

Measuring ultralow magnetic fields (picotesla), required for biomedical appliations or for space exploration, is challenging. Magnetoresistive sensors [1,2] are interesting candidates thanks to their compacity, low energy consumption and relatively low fabrication costs. Optimizing the detectivity of a magnetic sensor based on tunnel magnetoresistance (TMR) can be done by following two complementary options: i) improving the sensor sensitivity; ii) reducing the noise. Achieving these two objectives is not straightforward because there is a direct interplay between the sensitivity of a magnetic tunnel junction and its noise [3]. Indeed, the most prominent noise in magnetic tunnel junctions is nowadays the noise due to thermal fluctuations of the magnetization, which increases at low frequency and is called 1/f magnetic noise [1,4,5]. A given thermal fluctuation of the magnetization induces a fluctuation of the resistance that contributes to the noise, and that is also directly proportional to the sensitivity. Keeping in mind this limitation, the problem is nevertheless not hopeless: the noise can be reduced by increasing the total magnetic volume, i.e., the size and number of junctions. It is also possible to increase the sensor sensitivity by using a flux concentrator (FC) that amplifies the magnetic field at the junction site; this gain of sensitivity is obtained without changing the junction response, so the magnetic noise remains the same. Hence to aim at the picotesla range at low frequency, these sensors have been combined with magnetic FC of various sizes and shapes [5–10] that can increase the magnetic field at the sensor location by a factor up to 440 for single staged [11] and 520 for double staged FC [10].

In this paper, we choose to follow these two options simultaneously and estimate the improvement of detectivity that can be obtained by using a flux concentrator and by increasing the number and size of junctions. All the junctions must be packed within the air-gap of the flux concentrator, but more junctions require a wider air-gap, which in turn results in a reduction of the flux concentrator gain. The optimal trade-off must therefore be estimated by modeling. We first modeled the dependence of the concentrator gain on its geometry and tried to reproduce the obtained behavior with theory. Then we included this information together with the properties of the MTJs in an overall model of the sensor detectivity to determine the optimal sensor design.

We first performed finite elements simulations of the flux concentrator using Comsol Multiphysics® software in order to compute the dependence of the concentrator gain as a function of its geometrical parameters. The flux-concentrator is made of NiFe with a relative magnetic permeability $\mu_r$ = 1500 and a thickness t = 8 µm. The length and width of one arm of the flux concentrator are L = 5 mm and W = 2.3 mm with a wedge length $L_2$ of 2 µm (see Fig. 1.a).

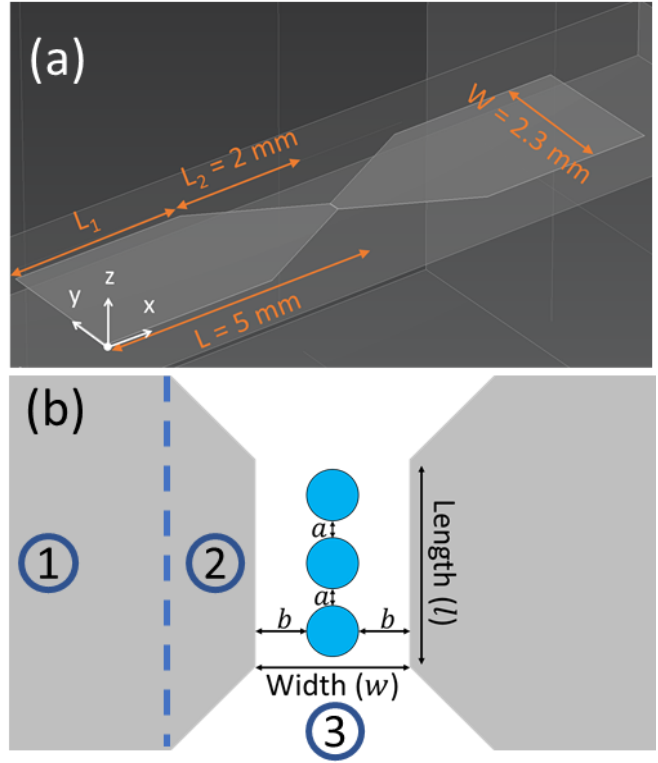

*Fig 1: (a) Design of the flux concentrators used in the Comsol Multiphysics® simulations, the thickness of the magnetic flux concentrators is 8 µm; (b) Scheme of the air-gap with MTJs represented as blue disks. The areas marked by a 1, 2 and 3 are the areas where the magnetic reluctance is computed later on.*

The flux concentrator gain is calculated as the ratio of the magnetic field in the middle of the air-gap with respect to the external magnetic field (1 µT) applied along the x direction. To properly describe how the flux concentrator traps the flux lines, a large volume is simulated: the flux-concentrator is placed in the middle of a large air box. The field inside the simulation boundaries is defined by analogy with an electrical potential, using a Coulombian description as explained in the work of X. Zhang et al [12]. The mesh is much finer close to the air-gap and particularly fine inside the air-gap in which each element size is between one and two µm.

Our work aims at optimizing the FC design in conjunction with the integration of the magnetic tunnel junctions, seeking a balance between the gain achieved by the design of the FC and the total volume of MTJs. The varying geometrical parameters of the concentrator are the width $w$ and length $l$ of its air-gap, which will accommodate a certain number $N$ of junctions as illustrated in Fig. 1.b. The "$a$" and "$b$" parameters shown in Fig. 1.b are spaces linked to lithography restrictions and will be used below in the optimization model, together with the number $N$ of junctions and their optimal diameter $d$.

Since an increase of the diameter of the MTJs directly results in a wider air-gap between the concentrators, we have first studied the variation of the gain for different air-gap widths, with an arbitrary fixed length of 200 µm. Figure 2 shows this variation along the x-axis (Fig. 2a) and the y-axis (Fig. 2b).

As expected, increasing the gap width drastically reduces the gain. For example, doubling it from 5 to 10 µm reduces the gain by almost half (from 495 to 258), and an increase beyond 15 µm reduces the gain below 200. On the other hand, it can be observed (Fig. 2b) that the gain is the same for a junction placed at the edge (in *y* direction) of the air-gap or in its center, as long as it remains within the air-gap. Indeed, due to the shape of the concentrator, the field lines are very tightly packed and there are no significant gain losses at the edges, while a drastic drop is observed outside the air-gap.

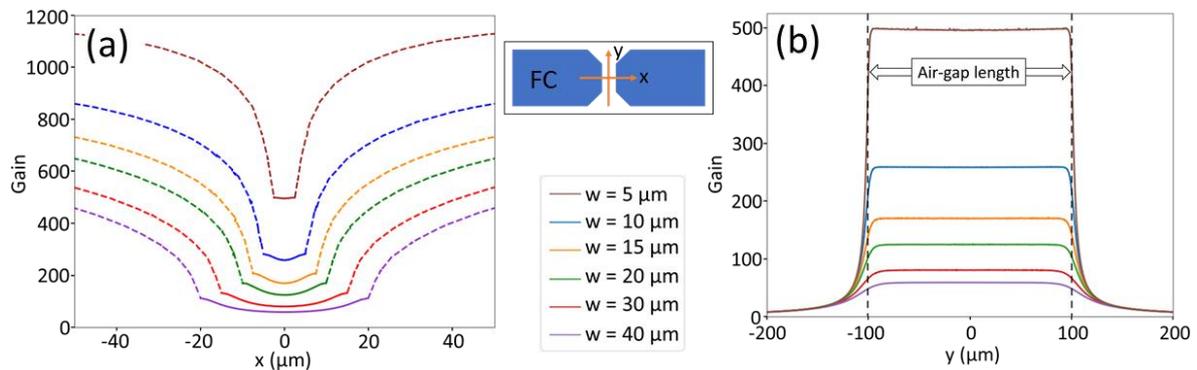

*Fig 2: Evolution of the magnetic gain as a function of the air-gap width w for a fixed length l of 200 µm: (a) Measured along the width (x axis) and in the middle of the length, the dashed part corresponds to the inside of the flux concentrator; (b) Measured along the length (y axis) and in the middle of the width. The black dashed lines mark the dimension of the air-gap along its length.*

It appears therefore that increasing the magnetic volume of the sensor does not have to be achieved solely by increasing the junction diameter, but rather by using more MTJs in the gap, arranged along y-axis and interconnected. This approach would prevent an excessive increase of the air-gap width, which is detrimental to the gain, as previously observed. Nonetheless, elongation reduces the concentration of field lines in the gap, thereby also reducing the gain. Fig. 3 shows a more complete view of the variations of the gain with the width and length of the air-gap. In Fig. 3a we observe that the gain strongly decreases when the air-gap width increases, following an expected $1/w$ law. The impact of the air-gap length on the gain is less critical: Fig. 3b shows a gentle decrease of the gain for increasing length. Fig. 3.c summarizes the simulated variation in gain as a color map with both parameters, clearly highlighting the strong dependence on the gap width and a much weaker variation with the gap length. We can easily conclude from Fig. 3 that it is more efficient to add MTJs in the length of the air-gap compared to the width.

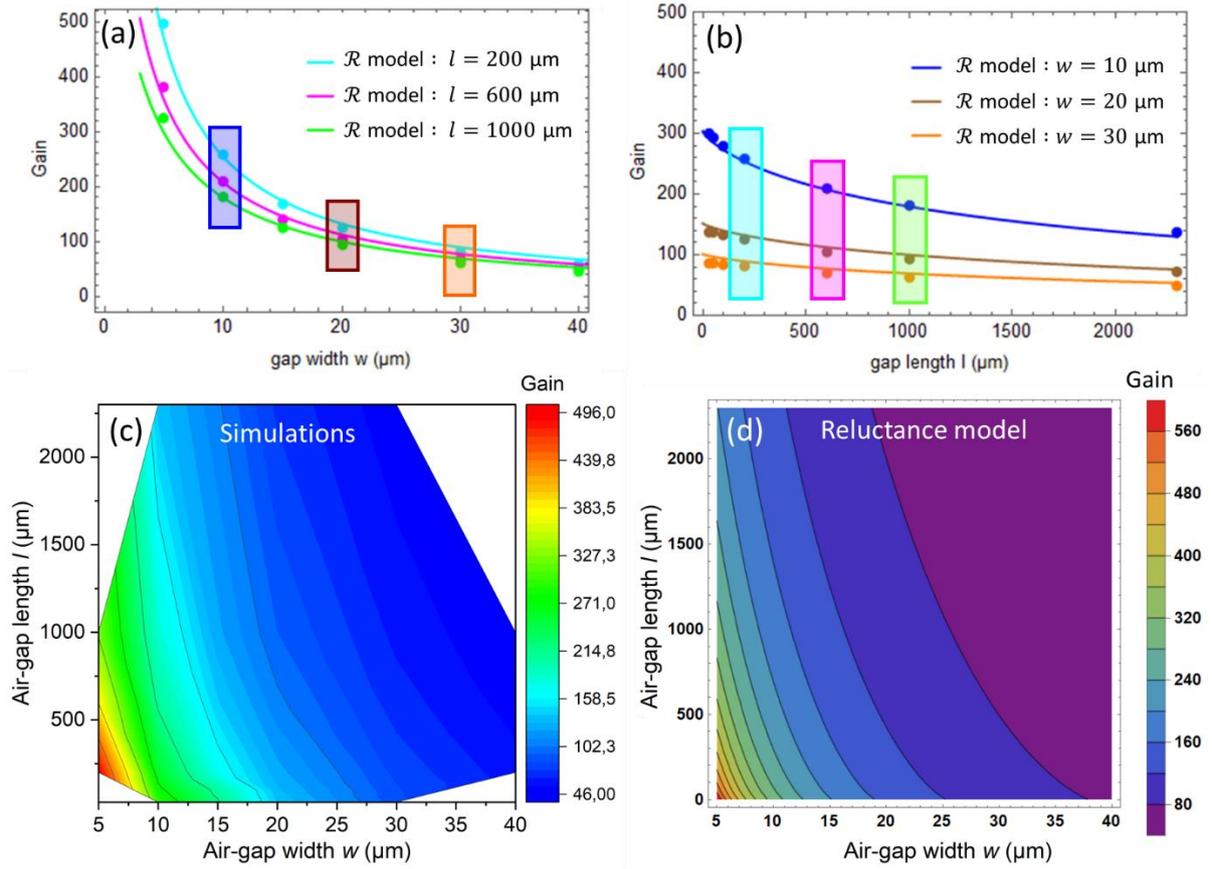

*Fig 3: Evolution of the magnetic gain as a function of, (a) the air-gap width w for three fixed lengths and (b) the air-gap length l for three fixed widths. The points show the simulated results and the solid lines are fits obtained from the reluctance model. The colored areas highlight the points in common between both plots. Color maps showing the influence of both the width and the length of the air-gap on the gain obtained from the simulation (c) and from the reluctance model (d).*

In order to understand and reproduce the observed behavior of the concentrator gain, we used the concept of magnetic reluctance $\mathcal{R}$ [12], which links the magnetic flux $\phi$ in a magnetic core to the current flowing in a coil around it ($U = nI$ with $n$ the number of spires and $I$ the current in the coil): $\mathcal{R}\phi = U$. Although magnetic reluctance is defined for closed magnetic circuits, in which the magnetic flux is conserved, we adapted this concept to model the gain of the flux-concentrator. By definition, similarly to an electrical resistance, the magnetic reluctance of a magnetic core of length $L$ and cross-section $S$ is $\mathcal{R} = \frac{1}{\mu}\frac{L}{S}$, where $\mu = \mu_0\mu_r$ is the magnetic permeability. In our case, the reluctance of the flux-concentrator is the sum of the reluctance of the two magnetic arms and the one of the air-gap since reluctances in series add like electrical resistances. Due to symmetry, calculating the reluctance of half the flux-concentrator is enough, the total reluctance being twice larger. The reluctance to be calculated is $\mathcal{R} = \mathcal{R}_1 + \mathcal{R}_2 + \mathcal{R}_3$, where $\mathcal{R}_1$ is the reluctance of the rectangular part of the magnetic arm, $\mathcal{R}_2$ is the reluctance of the wedged region and $\mathcal{R}_3$ the reluctance of the half air-gap, with the notations used in Fig. 1.b. We obtain:

$$\mathcal{R} = \frac{1}{\mu_0\mu_r}\left(\frac{L_1}{Wt}\right) + \frac{1}{\mu_0\mu_r}\left(\int_0^{L_2}\frac{dx}{t\left(W - \frac{W-l}{L_2}x\right)}\right) + \frac{1}{\mu_0}\left(\frac{w}{2lt}\right)$$

After integration, the result is:

$$\mu_0 t \, \mathcal{R} = \frac{1}{\mu_r}\left[\frac{L_1}{W} + \frac{L_2}{W-l}\ln\left(\frac{W}{l}\right)\right] + \frac{w}{2l}$$

Since the magnetic flux in the air-gap is $\phi = Blt = \frac{U}{\mathcal{R}}$, the field in the air-gap is $B = \frac{U}{\mathcal{R}lt}$. Therefore, the dimensionless gain $G$ is expected to be inversely proportional to $\mathcal{R}l$.

Finally, we obtain:

$$1/G \propto \frac{l}{t\mu_r}\left[\frac{L_1}{W} + \frac{L_2}{W-l}\ln\left(\frac{W}{l}\right)\right] + \frac{w}{2t}$$

This formula fits quite well the simulated gain as a function of the gap width. However, it does not fit accurately the simulation data as a function of the air-gap length $l$. This should not be entirely surprising, since using reluctance in a system with no flux conservation is a little far-stretched. However, the simulation data can be fitted by the formula providing that the length $L_1$ and $L_2$ are increased by a factor that scales with the air-gap width. We finally obtained a formula that fits all our sets of data, with no free parameter (except the $G_0$ prefactor set at 130), using the following simple modification of the above formula:

$$G(w,l) = G_0 \frac{\mathcal{R}_{mod}(L_1,L_2,W,w_0,W)}{\mathcal{R}_{mod}(L_1,L_2,W,w,l)} \text{ where } \mathcal{R}_{mod}(L_1,L_2,W,w,l) = \frac{l}{t\mu_r}\left(1+\frac{w}{10}\right)\left[\frac{L_1}{W}+\frac{L_2}{W-l}\ln\left(\frac{W}{l}\right)\right] + \frac{w}{2t}$$

In Fig. 3.a and b, all fits are performed with this formula, with a fixed (identical) value of the $G_0$ prefactor. The color map of the gain obtained by this reluctance model (Fig. 3d) show very similar trends as the one obtained by simulations (Fig. 3c). Even if the physical meaning of the modification of the reluctance formula has still to be fully understood, this analytic formula appears to be an interesting tool for designing flux concentrators and it proved to be useful for implementing our model to optimize the sensor detectivity. Further studies are nevertheless required in order to assess the robustness of this formula when geometrical parameters $L_1$, $L_2$ and $W$ are varied.

Now that we have characterized the dependence of the gain on the concentrator geometrical parameters, we can use this information to optimize the overall sensor design. The sensing elements to be placed within the air-gap are MTJs of circular shape connected in a serial/parallel configuration and placed on one single row (see Fig.1b). The number $N$ of junctions and their diameter $d$ are the parameters we wish to modify to improve the detectivity. Since the bottom electrode must be larger than the junction disk and the air-gap larger than the bottom electrode, the junction diameter $d$ has a direct influence on the air-gap width. This one must be determined by taking reasonable margins related to lithography reposition issues. In the model, we used $w = d + 2b$ with a global margin $b$ = 3 μm at both sides of the junction disk. The number $N$ of junctions will in turn influence the length of the air-gap. Etching issues require a minimal spacing of $a$ = 1.5 μm between disks. The air-gap minimal length is thus $l = Nd + (N-1)a = N(d+a) - a \approx N(d+a)$.

The detectivity (also called limit of detection) of the sensor is defined as the noise to sensitivity ratio. For a variation $\delta H$ of the external magnetic field, the resistance of the MTJ varies by $\delta R$, so the MTJ voltage variation is: $\delta V = \delta R \, I_0 = \frac{dR}{\mu_0 dH}\frac{V_0}{R}\mu_0 \delta H = s_0 V_0 \mu_0 \delta H$, where $s_0$ is the normalized ohmic sensitivity of the junction (expressed in Ω/T. Ω$^{-1}$) and $V_0$ the voltage bias. When the junction is placed within the flux concentrator air-gap, the final sensitivity $S = \delta V/(\mu_0 \delta H)$ expressed in V/T is thus equal to $Gs_0V_0$ or, with more usual units, to $10GsV_0$, with $G$ the flux concentrator gain, $s$ the normalized sensitivity (in %/mT, as commonly used in literature [1,13–15]) and $V_0$ the voltage bias. The factor 10 comes from unit conversions. On the other hand, the noise of the junction is the sum of the Johnson noise, the shot noise, the electric 1/f noise and the magnetic

1/f noise. Since the latter is the most prominent noise at low frequency in high quality MTJ [1,4,10], we consider only the output noise from magnetic origin whose power spectral density can be written as $S_V = \frac{\alpha_H V_0^2}{Af}$ (in V²/Hz), where $f$ is the frequency, $\alpha_H$ the Hooge parameter and $A$ the area of the junction. For a sensor with $N$ interconnected junctions, $A = N\frac{\pi d^2}{4}$. Thus, the sensor detectivity (in T/Hz$^{1/2}$) is:

$$D = \frac{\sqrt{S_V}}{S} = \frac{\sqrt{\alpha_H}}{\sqrt{Af}\, 10 Gs} = \frac{\sqrt{\alpha_H}}{10s} \frac{2}{\sqrt{\pi f}} \frac{1}{d\sqrt{N} G(w,l)} \quad (1)$$

with $w = d + 2b$ and $l = N(d + a)$

The optimization of the geometrical parameters requires to minimize the function $F(d, N)$ defined as:

$$F(d, N) = \frac{1}{d\sqrt{N} G(d + 2b, N(d + a))} \quad (2)$$

The total number of junctions is limited by the fact that the length of the junctions' row must be smaller than the width of the flux concentrator i.e., $N(d + a) \leq W$. This requires that $d \leq d_{max}(N)$ with $d_{max}(N) = \frac{W}{N} - a$. Fig. 4 shows a color map of the function $F(d, N)$ as a function of the number of junctions $N$ and the junction diameter $d$. In this map, the red line shows the maximal value allowed for the junction diameter $d_{max}$. Points beyond this line must therefore be discarded.

Along a horizontal line on this chart, the junction diameter is fixed and the number of junctions increases. Thus, the flux-concentrator air-gap width is fixed and only its length increases when the number of junctions increases, resulting in a slight reduction of the FC gain. By contrast, the noise of the sensor strongly decreases as $1/\sqrt{N}$. The net result is a larger signal to noise ratio thus a better detectivity (lower values of $F$ function).

When considering a vertical line on this chart, the number of junctions is fixed and their diameter increases. Therefore, the FC air-gap widens, which strongly reduces its gain. However, the detectivity varies like the inverse of the product of the diameter with the gain (see eq. (1)). To the first order, $d * G \propto \frac{d}{w}$. Since $w = d + 2b$, the function $\frac{d}{w}$ increases with $d$ (see inset of Fig. 4, plotted for $N = 44$ and two fixed values of $l$). However, when $d$ increases, $l$ must also increase to accommodate these larger junctions. This results in a non-monotonous dependence of $d * G$ with $d$ : it varies between the two dashed lines corresponding to the minimum length $l_0$ (for 44 junctions of diameter 5 μm) and maximum length of the air-gap ($W$). The position of the maximum of $d * G$ corresponds to the highest signal to noise ratio, thus the better detectivity. The overall optimum is obtained when this extremum is reached for a number of junctions such as the extremum coincides with the maximal diameter allowed.

We observe in Fig. 4 that the minimal value of $F$ is obtained close to the red line, thus for a rectangular flux concentrator without wedge. This result shows that the improvement of the gain due to the wedge does not compensate for the reduction of the junction's number. The minimum value of the function F is evaluated at 7.72 10$^{-5}$ μm$^{-1}$ and is obtained for $N$ between 160 and 166 and with a junction diameter slightly below 13 μm.

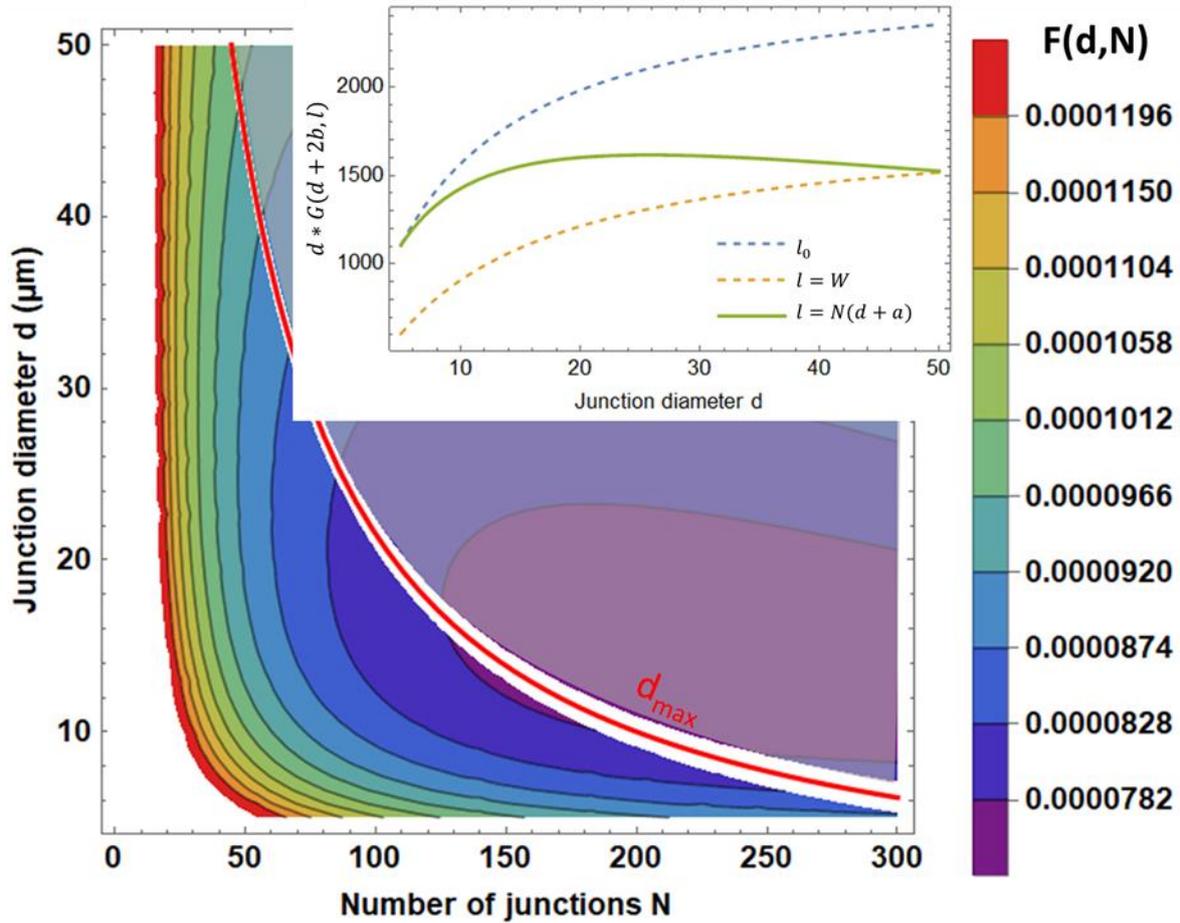

Fig. 4: Color map of the function $F$ ; the red line corresponds to the maximal diameter allowed for the row of junctions to fit within the length of the air-gap. Inset: dependence of the product $d*G(w,l)$ for 2 extreme values of $l$: $l = l_0$ and $l = W$ (dashed line) and for $l = N(d+a)$ (solid line); the values chosen here are $l_0 = 286$ μm and $N = 44$.

In order to calculate the detectivity from this value, we need the value of $\sqrt{\alpha_H}/(10s)$ where $\alpha_H$ is the Hooge factor and $s$ the sensitivity. We have previously mentioned that increasing the sensitivity should also increase the magnetic noise. Experimentally, this relationship was demonstrated in [3,16] who measured the detectivity as a function of the voltage bias on the junction. Both these studies found that the TMR ratio and thus the sensitivity decreased with increasing bias; at the same time, the noise was shown to follow the same trend so that the detectivity remained constant.

| reference | $\alpha_H$ (μm²) | $s$ (%/mT) | $\sqrt{\alpha_H}/(10s)$ ($10^{-6}$ T.μm) | Comment |
|---|---|---|---|---|
| [3] | $1.10^{-7}$ | 26 | 1.2 | $t_{Ru}$=0.3 nm |
| [3] | $1.10^{-7}$ | 5 | 6.3 | $t_{Ru}$=0 nm |
| [13] | $4.10^{-8}$ | 18 | 1.1 | |
| [16] | $8.10^{-8}$ | 3.7 | 7.6 | |

Table 1. Hooge parameter, sensitivity and $\sqrt{\alpha_H}/(10s)$ ratio extracted from various studies.

We can estimate the ratio $\sqrt{\alpha_H}/(10s)$ using the parameters extracted from previous studies. The obtained values are shown in Table 1 and range approximately from $1.10^{-6}$ to $7.10^{-6}$ T.μm. From these, let us assume a value $\frac{\sqrt{\alpha_H}}{10s} = 2.10^{-6}$ $T.\mu m$, which is reasonably optimistic and corresponds to a Hooge factor of $1.6\ 10^{-7}$ μm² for a junction with a sensitivity of 20%/mT. With this assumption, the detectivity at the optimum point would be 55 pT/$\sqrt{Hz}$ at 10 Hz. This value has to be compared with the detectivity of 55 nT/$\sqrt{Hz}$ for the same MTJ alone without flux-concentrator. Thus, the use of a flux concentrator and of a large number of junctions with the optimum design allows to gain 3 orders of magnitude on the sensor detectivity. For the optimal design found here, the flux-concentrator is a rectangular FC, without wedge, with an air-gap width of about $d + 2b = 19$ μ$m$ and a gain of 80.

To summarize, using finite element simulations and fitting a reluctance model, we modelled the gain of the flux concentrator with the air-gap dimensions that welcome a number of series/parallel magnetic tunnel junctions. Our model shows that the better detectivity is obtained for relatively small junctions and using a flux concentrator with no wedged region. This is explained by a larger reduction of noise with the magnetic volume that largely compensates the loss of gain of the FC while increasing its length. Our reluctance model of the FC gain is very satisfactory with our geometry and could be further adapted to study other FC geometries.

**Acknowledgements**

The authors thank Claude Cavoit for fruitful discussions. This work was supported by the CNES R&T program, the ANR-22-CE42-0020 project MAROT and by France 2030 government investment plan managed by the French National Research Agency under grant reference PEPR SPIN –ADAGE ANR-22-EXSP-0006.